\documentclass[aps,pra,reprint,superscriptaddress,longbibliography]{revtex4-2}
\usepackage[utf8]{inputenc}
\usepackage[english]{babel}
\usepackage[T1]{fontenc}
\usepackage{amssymb}
\usepackage{amsmath}
\usepackage{amsfonts} 
\usepackage{hyperref}
\usepackage{float}
\usepackage{enumitem}
\usepackage{tikz}
\usetikzlibrary{quantikz}
\usepackage{bbold}
\usepackage{natbib}
\usepackage{cancel}
\usepackage{comment}

\usepackage[normalem]{ulem}
\newtheorem{definition}{Definition}

\usepackage{lipsum}

\makeatletter
\def\@opargbegintheorem#1#2#3{\trivlist
   \item[]{\bfseries #1\ #2\ (#3)} \itshape}
\makeatother

\begin{document}
\definecolor{colorSQG}{RGB}{63,113,186}
\definecolor{colorHAM}{RGB}{222,179,54}  
\renewcommand{\figurename}{FIG.}

\title{Impact and mitigation of Hamiltonian characterization errors\\ in digital-analog quantum computation}
\author{Mikel Garcia-de-Andoin}\email{mikel.garciadeandoin@tecnalia.com}\affiliation{TECNALIA, Basque Research and Technology Alliance (BRTA), Astondo Bidea Edificio 700, 48160 Derio, Spain}\affiliation{Department of Physical Chemistry, University of the Basque Country UPV/EHU, Apartado 644, 48940 Leioa, Spain}\affiliation{EHU Quantum Center, University of the Basque Country UPV/EHU, Barrio Sarriena s/n, 48940 Leioa, Spain}
\author{Alatz \'Alvarez-Ahedo}\affiliation{Department of Physical Chemistry, University of the Basque Country UPV/EHU, Apartado 644, 48940 Leioa, Spain}
\author{Adrián Franco-Rubio}\affiliation{Max-Planck-Institut für Quantenoptik, Hans-Kopfermann-Str. 1, D-85748 Garching, Germany}\affiliation{Munich Center for Quantum Science and Technology (MCQST), Schellingstr. 4, D-80799 Munich, Germany}\affiliation{University of Vienna, Faculty of Physics, Boltzmanngasse 5, 1090 Vienna, Austria}
\author{Mikel Sanz}\affiliation{Department of Physical Chemistry, University of the Basque Country UPV/EHU, Apartado 644, 48940 Leioa, Spain}\affiliation{EHU Quantum Center, University of the Basque Country UPV/EHU, Barrio Sarriena s/n, 48940 Leioa, Spain}\affiliation{IKERBASQUE, Basque Foundation for Science, Plaza Euskadi 5, 48009 Bilbao, Spain}\affiliation{Basque Center for Applied Mathematics (BCAM), Alameda Mazarredo 14, 48009 Bilbao, Spain}

\date{\today}

\begin{abstract}
Digital-analog is a universal quantum computing paradigm which employs the natural entangling Hamiltonian of the system and single-qubit gates as resources. Here, we study the stability of these protocols against Hamiltonian characterization errors. For this, we bound the maximum separation between the target and the implemented Hamiltonians. Additionally, we obtain an upper bound for the deviation in the expected value of an observable. We further propose a protocol for mitigating calibration errors which resembles dynamical-decoupling techniques. These results open the possibility of scaling digital-analog to intermediate and large scale systems while having an estimation on the errors committed.
\end{abstract}

\maketitle

\section{\label{sec:intro}Introduction}

Quantum computing was first introduced as a way of simulating quantum systems with a controllable device \cite{Benioff1980, Feynman}. There are two main protocols to control these quantum systems. On the one hand, analog quantum computing (AQC) acts on the system by a continuous variation of some parameters \cite{Arnab2008}. Its main advantage is the robustness against noise at the cost of a limited set of control actions over the system. On the other hand, digital quantum computing (DQC) acts on a state with a series of discrete unitary evolutions or gates \cite{Deutsch1995}. A crucial advantage of DQC is the capacity of implementing a universal set of operations. This allows DQC to implement error correcting codes enabling it to perform simulations without errors \cite{aharonov1996, Kitaev1997, Bultrini2022}. However, the algorithms implemented in DQC are usually less resistant to noise compared to AQC \cite{Poggi2020, Altman2021, adrian2023, Fauseweh2024}. Even though we are closer to having access to hardware and techniques allowing this \cite{Bluvstein2024, Bravyi2024}, we are still in the noisy intermediate scale quantum devices (NISQ) era \cite{Preskill2018NISQ}. In this situation, there is a focus on noise mitigation techniques that allow running simulations with less errors \cite{Zhenyu2023}. This would help to fulfill the requirements of the quantum threshold theorem \cite{knill1996,Aharonov2008} and advance to the fault-tolerant era.

With this challenge in mind, digital-analog quantum computing (DAQC) \cite{Adrian2020DAQC} was proposed as a way of achieving the robustness of AQC while maintaining the flexibility of DQC. Instead of trying to mitigate the crosstalk of DQC systems, DAQC employs the natural interaction Hamiltonian as a resource to generate entanglement. Universality of DAQC is achieved by alternating the application of single qubit gates (digital blocks) with periods in which the system freely evolves under the system Hamiltonian for a given time (analog blocks) \cite{Dodd2002UnivQC}. It has been extensively shown that this new paradigm allows the implementation of quantum algorithms \cite{Ana2020,  Ana2023, jonathan2024, pasqal2024qadence}, allowing for efficient circuit designs. The results from \cite{garcia2022noise} showed a better performance of DAQC against noise compared to similar DQC circuits. Additionally, an error mitigation technique based on zero-noise extrapolation tailored for DAQC was proposed.

A great part of the quantum computing community has its focus on developing new tools to reduce the errors in the systems, in order to implement any operation. This is currently done by the engineering of quantum systems and their controls \cite{Arute2019, Conner2021, Kim2023} and with the development of active mitigation techniques \cite{Zhenyu2023} such as dynamical-decoupling techniques \cite{Souza2011, Ezzell2023}, zero noise extrapolation \cite{Giurgica2020}, or probabilistic error cancellation \cite{Temme2017, Berg2023}. The question of reducing the errors up to a threshold to make the quantum devices useful has led to the conclusion that errors must decrease exponentially with the size of the system to perform fault tolerant computations \cite{knill1996,Aharonov2008, Takagi2022}. However, another question can be asked about errors. Is it possible to have a quantum system with such properties that allow us to increase its size without jeopardizing its performance? This question can be answered by looking at the stability of quantum simulation tasks, which can be studied from the error of measuring a certain observable in the thermodynamic limit with constant errors with respect to the system size \cite{Aharonov2022, Watson2022, adrian2023, kashyap2024}.

In this article, we study the errors committed in DAQC circuits for quantum simulation tasks under calibration errors in the coupling constants. We first study the difference between the target Hamiltonian and the Hamiltonian simulated with the DAQC circuit. Then, we obtain a bound for the error in the expectation value of an observable. From the insights obtained from these calculations, we develop a new DAQC circuit synthesis method aimed at reducing the effect of the calibration errors. For this, we trade off the total time of the circuit for the stability of the circuit.

The rest of the article is organized as it follows. Sec.~\ref{sec:brief} introduces the basis for DAQC. In Sec.~\ref{sec:stability} we study in detail the stability of DAQC, and Sec.~\ref{sec:mitigation} follows with the introduction of the error mitigation technique to enhance the stability. In Sec.~\ref{sec:discussion} we conclude by discussing the applicability of our results to particular experimental setups and we summarize our results in Sec.~\ref{sec:conclusions}.

\section{Brief introduction to DAQC}\label{sec:brief}

\begin{figure}
    \centering
    \includegraphics[width=1.0\linewidth,trim={10pt 0 0 0},clip]{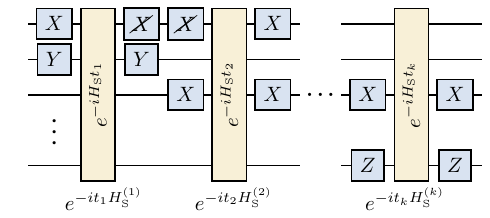}
    \caption{Representation of a generic stepwise-DAQC circuit. These circuits are composed of single qubit gates (blue) that act in between the analog blocks (yellow), which correspond to the free evolution of the system under the natural interaction Hamiltonian for a set time $t_k$. The sandwiching of SQGs changes the effective sign of the analog Hamiltonian for each block, $e^{-it_kH_\text{S}^{(k)}}$. Note that in stepwise-DAQC, the interaction Hamiltonian is turned off during the application of the SQGs. In contrast, in banged-DAQC the interaction Hamiltonian is always on while the gates are applied.}
    \label{fig:generalDAQC}
\end{figure}

In general, the goal of a DAQC paradigm is to construct the unitary generated by the evolution under a given problem Hamiltonian $H_\text{P}$ for a time $T$, $U=e^{-iTH_\text{P}}$. The problem Hamiltonian may contain arbitrary single-body terms and two-body interactions. As single-body terms can be implemented through single-qubit gates, our main focus is on the two-body terms ${H_\text{P} = \sum_{\mu,\nu\in\{x,y,z\}}\sum_{i<j}^N{h_\text{P}}_{ij}^{\mu\nu}\sigma_i^\mu\sigma_j^\nu}$. However, a given platform might provide us with a different source Hamiltonian ${H_\text{S} = \sum_{\mu,\nu\in\{x,y,z\}}\sum_{i<j}^N{h_\text{S}}_{ij}^{\mu\nu}\sigma_i^\mu\sigma_j^\nu}$. The task consists in the implementation of $U$ by the sequential application of digital-blocks composed of single-qubit gates (SQG) and analog-blocks of free evolution under the source Hamiltonian $H_\text{S}$ for a given prescription of times. For the sake of simplicity, if we assign each Hamiltonian a graph where each qubit is a vertex and the edges are the nonzero couplings of the Hamiltonian labeled by the interaction axes, we will assume that the connectivity graph of the source Hamiltonian must cover the graph of the problem Hamiltonian. Indeed, let $\mathcal{S}$ be the graph associated with $H_\text{S}$ and $\mathcal{P}$ to $H_\text{P}$. Then, $H_\text{P}$ is directly simulatable if $\mathcal{P}\subseteq\mathcal{S}$. Otherwise, the graph can be decomposed into several sub-graphs which can be simulated in sequence of alternate SWAP blocks \cite{Galicia2020EnhancedConnect}. To obtain a correct circuit, we make use of the following correspondence, $\sigma_\nu e^{-iH}\sigma_\nu^{\dagger}=e^{-i \sigma_\nu H \sigma_\nu^{\dagger}}$ for $\sigma_\nu$ a Pauli matrix. If we use the Pauli matrices plus the identity matrix to expand the Hamiltonian and as the applied SQGs, we find that this sandwiching structure allows us to change the signs of some of the terms of $H_\text{S}$, generating what we call an effective Hamiltonian as depicted in Fig.~\ref{fig:generalDAQC}. For simplicity in this paper, we will work in the stepwise-DAQC regime, in which $H_\text{S}$ is turned off during the digital blocks. 

Let us now introduce some notation. First, noticing that the only change in the effective Hamiltonian is a sign, we denote by $M_{ij\mu\nu,k}$ the sign of the term $\sigma_i^\mu\sigma_j^\nu$ after applying the prescription of SQGs corresponding to the $k^\text{th}$ analog block. Similarly, we denote by $t_k$ the prescribed time of evolution for the $k^\text{th}$ analog block. Now, the problem we want to solve is
\begin{equation}\label{eq:basicDAQC}
    e^{-i TH_\text{P}}=\prod_k e^{-i t_k \sum_{i,j,\mu,\nu}M_{ij\mu\nu,k}{h_\text{S}}^{\mu\nu}_{ij}\sigma^\mu_i\sigma^\nu_j}.
\end{equation}
By employing the first order of the Trotter-Suzuki expansion at the right-hand side of the expression, we obtain 
\begin{equation}\label{eq:approxDAQC}
    e^{-i TH_\text{P}}\approx \left(e^{-i\sum_k \frac{t_k}{q} \sum_{i,j,\mu,\nu}M_{ij\mu\nu,k}{h_\text{S} }^{\mu\nu}_{ij}\sigma^\mu_i\sigma^\nu_j}\right)^q,
\end{equation}
whose error can be reduced by increasing the number of Trotter steps $q$. Throughout this work, we assume that the Trotter error in this equation is negligible compared to the other errors studied here, $T\lVert H_\text{S}\rVert_\text{op}/q\ll1$ at the first order approximation. This expression allows us to solve this equation only by solving the exponent term by term. Thus, the design of a DAQC circuit consists of solving the following linear system of equations
\begin{equation}\label{eq:originalLSE}
    M\, t = T\, h_\text{P}\oslash h_\text{S},
\end{equation}
where $\oslash$ is the Hadamard or element-wise division \footnote{For two arrays of the same length and components $a=\{a_i\}$ and $b=\{b_i\}$, respectively, their Hadamard division is another array with the same length defined as $a\oslash b=\{a_i/b_i\}$.} and $h_\text{P}$ and $h_\text{S}$ are the arrays of couplings corresponding to $H_\text{P}$ and $H_\text{S}$, respectively, vectorized following a mapping $\alpha\leftrightarrow\{i,j,\mu,\nu\}$, $M_{\alpha, k}$ is the matrix of signs previously defined following the same mapping for $\alpha$ and $t$ is a column vector of times for the analog blocks which we want to compute. As the solution might not be unique, one can choose the one that minimizes the total time of the circuit $t_\text{A}=\lVert t\rVert_1$ using various techniques~\cite{MikelAlvaro2024,Bassler2024}. It was conjectured that the optimal total time could depend at most linearly with the system size $N$, $t_\text{A}=\mathcal{O}(N)$ \cite{Bassler2024TimeOptimal}. Prescriptions for SQGs leading to solvable systems of equations can be found in Ref.~\cite{MikelAlvaro2024}. More details about the definitions are given in Apx.~\ref{apx:definition}.

\section{Stability of DAQC under characterization errors}\label{sec:stability}

As discussed in the previous section, it is possible to simulate a target unitary $U=e^{-iTH_\text{P}}$ corresponding to a two-body problem Hamiltonian $H_\text{P}$ \cite{Dodd2002UnivQC} using a DAQC circuit based on the source Hamiltonian $H_\text{S}$ and SQGs. The robustness of this protocol against various coherent and incoherent errors has been extensively analyzed and compared with standard digital constructions \cite{Ana2020,garcia2022noise}. However, the stability of DAQC protocols in the presence of system calibration defects or unwanted interactions—i.e. inaccuracies in the determination of $H_\text{S}$—and their impact on protocol performance when scaling up remain open questions. In this section, we first investigate the effect of such defects and subsequently relate the findings to the stability of DAQC protocols.

Assume that calibration defects generate the Hamiltonian $H_\delta=\sum_{\{\{i,j\},\{\mu,\nu\}\}\in \mathcal{D}}{h_\delta}_{i,j}^{\mu,\nu}\sigma^\mu_i\sigma^\nu_j$, where $\lvert{h_\delta}_{i,j}^{\mu,\nu}\rvert\leq\delta$ and $\mathcal{S}\subseteq\mathcal{D}$. Additionally, we assume that any temporal component of the calibration error has a frequency much lower than $1/t_\text{A}$. This implies that the calibration error can be modeled as a constant deviation from the measured coupling during a circuit run. Thus, the real system Hamiltonian can be written as the sum of the measured and defect Hamiltonians, $H_\text{S}'=H_\text{S}+H_\delta$. 

\begin{figure}[t]
    \centering
    \includegraphics[width=\linewidth]{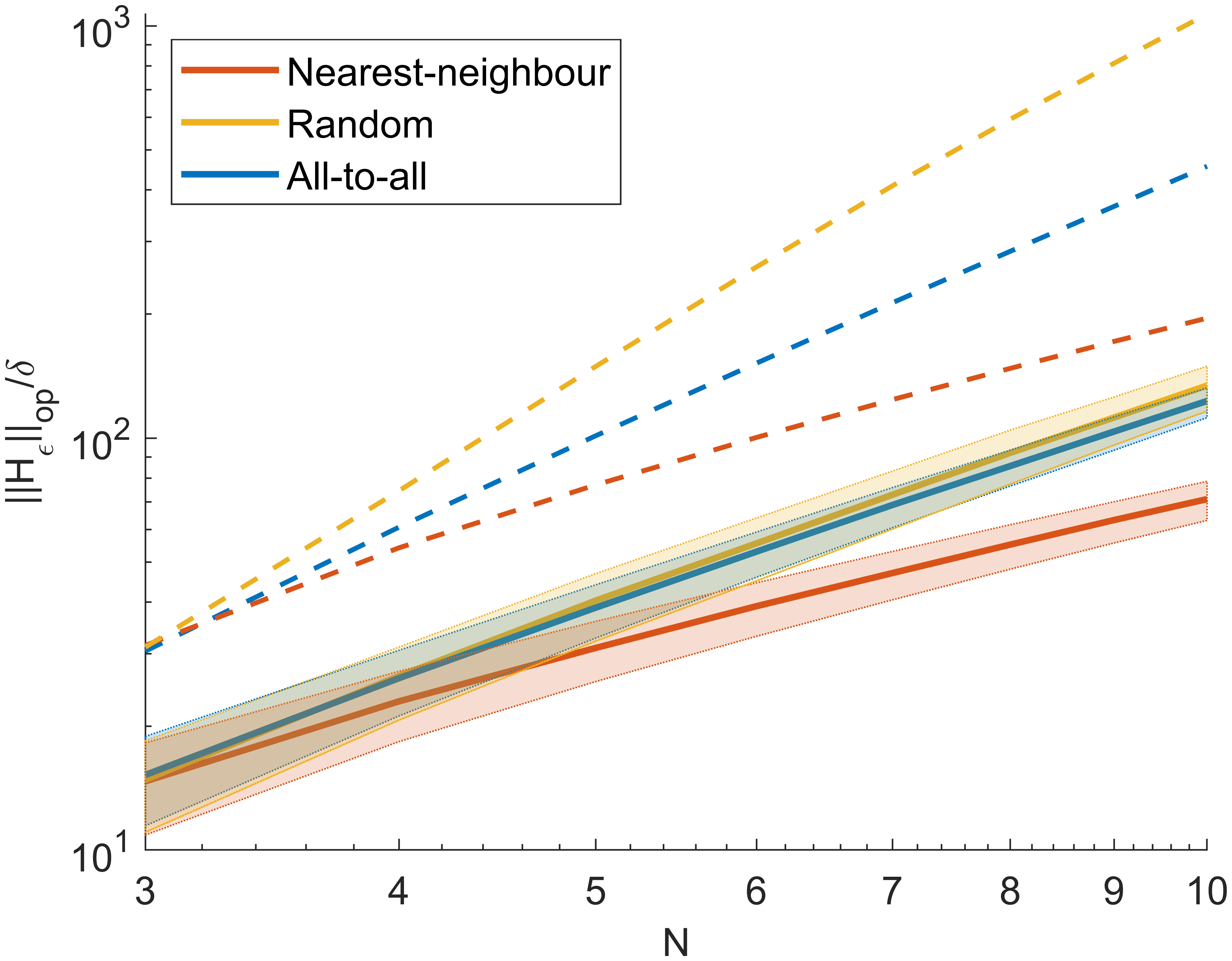}
    \caption{Operator norm of the error Hamiltonian for different system sizes and topologies: nearest-neighbours (red), random connected graphs (yellow) and all-to-all (ATA). Here $\mathcal{S}=\mathcal{P}$. The solid line represents the mean over 10000 different problems employing the usual DAQC protocol and the colored area the interquartile range. The dashed line shows the mean value of the bound given in Eq.~\ref{eq:errorHamiltonian}. More details about the simulations are given in the supplementary material.}
    \label{fig:bound}
\end{figure}

When calculating the DAQC schedule for a specific problem, $H_\text{S}$ is assumed to be correct. Consequently, the analog block times $t$ are determined as in Eq.~\ref{eq:originalLSE}. However, when the circuit runs with the actual system Hamiltonian $H_\text{S}'$, a systematic error appears in the simulated Hamiltonian, $H_\text{P}'=H_\text{P}+H_\varepsilon$. In the vectorized representation of the Hamiltonian, the error is bounded by the vector $p$-norm \footnote{The $p$-norm for a vector $x\in\mathbb{R}^N$ is defined as ${\lVert x\rVert_p=\left(\sum_{i=1}^N\lvert x_i\rvert^p\right)^{1/p}}$. Even though it does not fulfill the requirements of a norm, we will also employ the usual $-\infty$-norm notation to denote the minimum absolute value of any element in a vector.},
\begin{equation}\label{eq:errorvector} \lVert h_\varepsilon\rVert_p \leq\delta\lVert h_\text{P}\oslash h_\text{S}\rVert_p\big\rvert_\mathcal{S}+\delta \dfrac{t_\text{A}}{T} \lvert E_{\mathcal{D}\backslash\mathcal{S}}\rvert^{1/p}, \end{equation}
where $\lvert E_{\mathcal{D}\backslash\mathcal{S}}\rvert$ represents the number of edges in $\mathcal{D}$ which are not in $\mathcal{S}$, and $t_\text{A}=\lVert t\rVert_1$ is the total analog-block time. In Ref.~\cite{Bassler2024}, $t_\text{A}$ was upper bounded showing a dependency on the target evolution time $T$ and $\lVert h_\text{P}\oslash h_\text{S}\rVert_1$. In general, this value is not expected to grow faster than linearly with the number of qubits $N$~\cite{Bassler2024TimeOptimal}. By employing the bound on $h_\varepsilon$, we also express the error in terms of the operator norm \footnote{The operator norm is defined as the largest magnitude of the Hamiltonian's eigenvalues: $\lVert H\rVert_\text{op} = \sup {\lvert\lambda\rvert: H\ket{\psi}=\lambda\ket{\psi}, \lvert\braket{\psi}{\psi}\rvert^2=1}$.},
\begin{equation}\label{eq:errorHamiltonian} \lVert H_\varepsilon\rVert_\text{op} \leq\delta\lVert h_\text{P}\oslash h_\text{S}\rVert_1\big\rvert_\mathcal{S}+\delta \dfrac{t_\text{A}}{T} \lvert E_{\mathcal{D}\backslash\mathcal{S}}\rvert. \end{equation}
In Fig.~\ref{fig:bound}, we demonstrate the validity of this expression by numerically verifying various random problems solved using standard DAQC protocols. Proofs for Eqs.~\ref{eq:errorvector} and \ref{eq:errorHamiltonian} are provided in Apx.~\ref{apx:pnormvectorcouplings} and \ref{apx:operatornorm}, respectively.

With these results as a starting point, we now build the argument for the stability of DAQC. There are multiple manners to define the stability of a Hamiltonian evolution/simulation. Various definitions involve measuring the trace distance between the target state $\rho$ and the state obtained after the faulty evolution $\rho'$, given by $\text{tr}(\lvert \rho-\rho'\rvert)/2$ \cite{NielsenChuang2010, Suguru2019mitigating, zhuk2024trotter}. Some other approaches focus on the differences in the energy spectra of the target and simulated Hamiltonians \cite{Cubitt2018,zhou2021stronglyuniversal}. Other methods consider deviations in measured observables \cite{adrian2023} or use the distance between the unitary evolution matrices, as done in the analysis of Trotter-Suzuki formulas \cite{Suzuki1976Trotter}. Some of these definitions are complementary, meaning that if a process is stable under one definition, it will often be stable under other as well.

\subsection{Stability of DAQC simulations}
Here, in order to study the stability of DAQC, we will use a definition inspired by the spectral similarity,  similar to the one proposed in Ref.~\cite{Cubitt2018} but with the Frobenius norm, which imposes a strongest condition on the spectral similarity than the operator norm:

\begin{definition}[Stability of a quantum simulation]\label{def:stableEvolution}
Assume we aim to simulate the dynamics associated with an $N$-qubit target Hamiltonian $H_\text{P}$ for a time $T$. Consider that the DAQC protocol is designed using a source Hamiltonian $H_\text{S}$, but the actual source Hamiltonian is $H'_\text{S}=H_\text{S}+H_\delta$. Consequently, the effective target Hamiltonian produced deviates from the intended one, resulting in $H'_\text{P}=H_\text{P}+H_\varepsilon$. We define the protocol as stable under characterization errors if the deviation $H_\varepsilon$ from the target Hamiltonian is bounded by
    \begin{equation}\label{eq:condStableEvolution}
        \lVert H_\varepsilon \rVert_\text{F}\leq f(T,N)\,\lVert H_\delta \rVert_\text{F},
    \end{equation} 
    where $f$ is at most a polynomial function on T and $N$, and $\lVert \cdot \rVert_\text{F}$ is the Frobenius norm \footnote{The Frobenius norm of a Hamiltonian is defined as $\lVert H\rVert_\text{F}=\sqrt{\text{Tr}(H^\dagger H)}$}.
\end{definition}
This definition ensures that, in stable systems, the error caused by using a Hamiltonian with characterization inaccuracies will grow at most polynomialy with the system size.

Evaluating this definition, we confirm that DAQC fulfills the stability condition, as, for any protocol, we obtain
\begin{equation}
    \lVert H_\varepsilon\rVert_\text{F}\leq \left(\lVert h_\text{p} \oslash h_\text{s} \rVert_2^2 \big\rvert_\mathcal{S}+
    \lvert E_{\mathcal{D}/\mathcal{S}}\rvert \left(\frac{t_\text{A}}{T}\right)^2 \right)^{1/2}\lVert H_\delta\rVert_\text{F}.
\end{equation}
Here, it is easy to see that each of the two terms within the square root grow at most polynomically with $N$ \footnote{Although $t_\text{A}$ has no upper limit, the aforementioned conjecture states that optimized DAQC protocols' total time do not exceed the linear dependence on $n$.}. Calculations for this expression are given in Apx.~\ref{apx:frobeniusnorm}.

If we otherwise employ the operator norm as a way of evaluating Eq.~\ref{eq:condStableEvolution}, we can use Eq.~\ref{eq:errorHamiltonian} to study the stability under a modified version of Def.~\ref{def:stableEvolution}. Since $\lVert h_\text{P}\oslash h_\text{S}\rVert_1\leq N^2\lVert h_\text{P}\oslash h_\text{S}\rVert_\infty$, we can bound the operator norm of the error Hamiltonian by $\lVert H_\varepsilon\rVert_\text{op}\leq\left(\lVert h_\text{P}\oslash h_\text{S}\rVert_1\big\rvert_\mathcal{S}+\dfrac{t_\text{A}}{T} \lvert E_{\mathcal{D}\backslash\mathcal{S}}\rvert\right)\lVert H_\delta\rVert_\text{op}$, thus obtaining that DAQC is also stable employing the operator norm. Calculations are shown in Apx.~\ref{apx:operatornorm}.

\subsection{Error in the expectation value}

The stability discussed in the previous subsection is useful to obtain an upper bound for the error committed in problems where the task is to prepare a quantum state. However, in most cases, the task we want to perform does not require retrieval of tomographic information about the final state, but the expectation value of some observable. To study this, we will get inspiration from the definition provided in Ref.~\cite{adrian2023}:

\begin{definition}[Stable expectation value \cite{adrian2023}]\label{def:stability}
Let $\mathcal{L}$ be a target Lindbladian on an $N$-spin system. The (approximated) simulation of this quantum dynamics is stable with respect to an observable $\mathcal{O}$ if the difference between its expected value in a target state $\rho=\exp(t\mathcal{L})(\rho_0)$ and in the state prepared by the simulator $\rho'=\exp(t\mathcal{L}')(\rho_0)$
satisfies
\begin{equation}
    \Delta_\mathcal{O}=\lvert \mathrm{Tr}(\mathcal{O}\rho)-\mathrm{Tr}(\mathcal{O}\rho')\rvert\leq f(\delta),
\end{equation}
for a function $f$ dependent on an error parameter $\delta$ but independent of $N$, such that it goes to 0 when the error goes as well to 0, $f(\delta\rightarrow 0)\rightarrow 0$.
\end{definition}
In other words, an approximated dynamics is considered stable if the error in the estimation of an observable does not scale with the system size. While this general definition applies to spins of any dimension, in this work, we focus on qubit systems. Additionally, even though this definition allows for an arbitrary Lindbladian-type evolution, as in this work we only focus on coherent errors, we can substitute it by the usual Schrödinger-type evolution. It is noteworthy that, under this definition, stability strongly depends on the observable measured, particularly its support, $\text{supp}(\mathcal{O})$. Indeed, it is expected that observables involving a large subset of spins in the system induce a higher dependency of the error on the system size. A similar definition involving $N$ 1-local observables was used in Ref.~\cite{flannigan2022propagation}.

As this definition of stability heavily depends on the observable, we focus on bounding the error in the expectation value of an observable, $\Delta_\mathcal{O}$. Additionally, there is a dependence on the operator norm of the error Hamiltonian, $H_\varepsilon$, which can introduce a dependence on the system size. To avoid it, we can work on the small defect range, $\lVert H_\delta\rVert_\text{op}\ll\lVert H_\text{S}\rVert_\text{op}$. In this range of values and for sufficiently short times (see Apx.~\ref{apx:stability}), we can upper bound the error of the DAQC protocol by
\begin{multline}\label{eq:stabilityDAQC}
    \Delta_\mathcal{O}\leq 6\,T\,\delta\,\text{supp}(\mathcal{O})\text{deg}(\mathcal{P})\lVert\mathcal{O}\rVert_\text{op}\lVert h_\text{P}\oslash h_\text{S}\rVert_\infty\\
    +6\,t_\text{A}\,\delta\,\text{supp}(\mathcal{O})\text{deg}(\mathcal{D}\backslash\mathcal{S})\lVert\mathcal{O}\rVert_\text{op}.
\end{multline}
The first term of this expression accounts for errors in the calibration of the couplings associated with the connectivity graph $\mathcal{P}$. The second term corresponds to couplings initially assumed to be zero, $\mathcal{D}\backslash\mathcal{S}$. Since these couplings were not considered when designing the DAQC protocol, they may remain active throughout the circuit's duration without undergoing any effective sign change. Also note that as $t_\text{A}$ might grow with the system size, the stability could depend also on the problem characteristics. In Sec.~\ref{sec:mitigation} we discuss a error mitigation protocol that allow us to eliminate the dependence on the second term.

Both terms exhibit a dependence on the degrees of the graphs $\mathcal{P}$ and $\mathcal{D}\backslash\mathcal{S}$. In general, the degrees of these graphs could grow up to $N-1$. In other words, in order to have an error $\Delta_\mathcal{O}$ that is independent of the system size, it is necessary, but not sufficient, to have the associated graphs to have bounded degree. This case holds for systems where qubit couplings are designable or interactions depend on particle distances and arrangements, as discussed in Sec.~\ref{sec:discussion}. For geometrical local Hamiltonians—i.e. whose associated connectivity graphs are lattices—we can connect our results with the findings of Ref.~\cite{adrian2023}, where geometrical locality of spin interactions was required to prove stability. A proof for Eq.~\ref{eq:stabilityDAQC} is provided in Apx.~\ref{apx:stability}.

As an additional note, let us rephrase the error in the expectation value problem through a point of view closer to an experiment design. Assume that we want to know which is the maximum allowed calibration error $\delta$ given a desired maximum error $\Delta_\text{max}=\lvert \text{Tr}(\mathcal{O}\rho)-\text{Tr}(\mathcal{O}\rho')\rvert_\text{max}$. Using Eq.~\ref{eq:stabilityDAQC}, we find that the maximum allowed calibration error fulfilling the previous condition is upper bounded by $\delta\leq\Delta_\text{max}/(6\,\text{supp}(\mathcal{O})\lVert\mathcal{O}\rVert_\text{op}(T\,\text{deg}(\mathcal{P})\lVert h_\text{P}\oslash h_\text{S}\rVert_\infty+t_\text{A}\,\text{deg}(\mathcal{D}\backslash\mathcal{S})))$. This calculation helps us anticipate whether a given system fulfills the requirements for solving a given problem with a desirable error level.

\subsection{Particular examples}

As we have seen, stability can be studied directly for a generic simulation problem. Let us illustrate how we can estimate the error in expectation values for particular problems. Let us assume a ZZ-Ising spin system with 1D nearest-neighbour connectivity for both the hardware and problem Hamiltonians, so the associated graphs are path graphs ($\mathcal{P}=\mathcal{S}=P_n$). We further assume that the qubits interact up to second neighbours, but this long-range interaction is not taken into account for the characterization of $H_\text{S}$, and thus $\text{deg}(\mathcal{D}\backslash\mathcal{S})=2$. For this particular setup, the optimal total time is exactly $t_\text{A}=T\lVert h_\text{P}\oslash h_\text{S}\rVert_\infty$. The expression for the stability in this case can be bounded by $\Delta_\mathcal{O}\leq 6\delta\,T\,(\text{supp}(\mathcal{O})+1)\lVert h_\text{P}\oslash h_\text{S}\rVert_\infty $, which can only depend on the system size through the support of $\mathcal{O}$. 

Let us discuss another example in which we have a ZZ-Ising spin system but in a square lattice with periodic boundary conditions. Further assume that both the system and the problem are translationally invariant. For this, we consider a system in which the couplings between sites in the same row ($i,j\leftrightarrow i\pm1,j\rightsquigarrow h_\text{S,r},h_\text{P,r}$) have the same coupling constant ($h_\text{S,r}$,$h_\text{P,r}$), idem for the couplings within the same column ($i,j\leftrightarrow i,j\pm1\rightsquigarrow h_\text{S,c},h_\text{P,c}$). A general (but not unique) DAQC protocol for this case consists of 4 different digital-analog blocks: no gates applied, $X$ gates applied on all qubits within an even numbered column, $X$ gates applied on all qubits within an even numbered row, and $X$ gates applied to qubits whose column and row indices have different parities. Let us assume that the calibration errors can occur up to second neighbours. In this example, the expression for the error in the estimation of a local observable in a single qubit is $\Delta_\mathcal{O}\leq 24\,T\,\delta\,(\lvert h_\text{P,r}/h_\text{S,r}\rvert+\lvert h_\text{P,c}/h_\text{S,c}\rvert)\lVert\mathcal{O}\rVert_\text{op}$.

\section{Mitigation of calibration errors in DAQC schedules}\label{sec:mitigation}

In previous works \cite{Adrian2020DAQC,Galicia2020EnhancedConnect,Ana2023,MikelAlvaro2024,Ana2020,Bassler2023,Bassler2024}, when designing the DAQC schedules, the objective was to minimize the total annealing time. This was achieved by solving the linear system of equations from Eq.~\ref{eq:originalLSE} with a classical optimizer such that the solution $t$ had the minimal 1-norm. A strategy to further reduce $t_\text{A}$ involves removing all equations from Eq.~\ref{eq:originalLSE} that had an indeterminate form in the right-hand side, this is, we remove all rows from the problem which corresponded to a coupling constant of ${h_\text{S}}_{ij}^{\mu\nu}=0$. This effectively increases the number of free parameters of the system, allowing for solutions with lower $t_\text{A}$. This general strategy for optimizing the total time does not take into account the couplings assumed to be 0, as any selection of $t_i$'s satisfies the equation associated with these couplings. However, this indeterminate form is the source of the second term in the expression of the errors in Eqs.~\ref{eq:errorvector}, \ref{eq:errorHamiltonian} and \ref{eq:stabilityDAQC}.

Taking this into account, we use this indeterminate form to reduce the errors coming from the calibration defects of the system. To this end, we force all couplings that can potentially have a significant calibration defect—i.e., the couplings in $\mathcal{D}\backslash{S}$—to change their effective signs such that the simulated Hamiltonian ${h_\text{P}}_{ij}^{\mu\nu}$ is 0 regardless of the real coupling of the system ${h_\delta}_{ij}^{\mu\nu}$. This is equivalent to selecting 0 as the value for the indeterminate forms $(0/0\rightarrow0)$ that appear in the right-hand side of Eq.~\ref{eq:originalLSE}. Using this design choice, the expectation value error turns into
\begin{equation}\label{eq:stabilityErrorCorrected}
    \Delta_\mathcal{O}\leq 6\,T\,\delta\,\text{supp}(\mathcal{O})\,\text{deg}(\mathcal{P})\lVert\mathcal{O}\rVert_\text{op}\lVert h_\text{P}\oslash h_\text{S}\rVert_\infty.
\end{equation}
Here we have an even weaker sense of dependence on the system size, as there is no dependence on the total analog time. A downside of this strategy is an increase in the total time of the circuit compared to the original strategy, ${t_\text{A}}_\text{remove}\leq{t_\text{A}}_{0'\text{s}}$. As we are adding extra equations, we are reducing the number of free parameters of the system, and thus we are obtaining solutions with larger 1-norms. Consequently, there is a trade-off between the total analog time and the mitigation of the calibration error, which should be addressed individually for different cases. See Apx.~\ref{apx:totalTime} for an in-depth discussion about the effect on the strategy on the total time.

As an illustrative example, let us show the details of the application of this error mitigation technique in the smallest non-trivial example. Assume we have a ZZ-Ising spin system with 3 qubits but we only have a non-zero measurement for 2 of the couplings, $H_\text{S}={h_\text{S}}_{12}\sigma_1^z\sigma_2^z+{h_\text{S}}_{13}\sigma_1^z\sigma_3^z$. Assume also that the problems that we wish to solve have a maximum 2 non-zero couplings. We perform any simulation while employing the error mitigation protocol, we assign 0 to the missing coupling. Then, we have to solve the following problem
{\small
\begin{equation}
    \min\,\lVert t\rVert_1, \begin{pmatrix}
        1 & -1 & -1 & 1\\
        1 & -1 & 1 & -1\\
        1 & 1 & -1 & -1
    \end{pmatrix}\begin{pmatrix}
        t_1\\
        t_2\\
        t_3\\
        t_4
    \end{pmatrix}=T\begin{pmatrix}
        {h_\text{P}}_{12}/{h_\text{P}}_{12}\\
        {h_\text{P}}_{13}/{h_\text{P}}_{13}\\
        0
    \end{pmatrix}, t_i\geq0.
\end{equation}}
To check the resilience against errors, we can evaluate the stability condition. Assume that we have a maximum characterization error of $\delta$, such that the Hamiltonian of the error is $H_\delta=\delta_{12}\sigma_1^z\sigma_2^z+\delta_{13}\sigma_1^z\sigma_3^z+\delta_{23}\sigma_2^z\sigma_3^z$. The Frobenius norm of the deviation Hamiltonian, $H_\varepsilon$, is $\lVert H_\varepsilon\rVert_\text{F}=T\sqrt{\lVert h_\text{P}\oslash h_\text{S}\rVert_\infty^2+\lVert h_\text{P}\oslash h_\text{S}\rVert_{-\infty}^2} \lVert H_\delta\rVert_\text{F}$. In this case, we see that the only dependence on the number of qubits is through the square root of the number of couplings, and thus, following Def.~\ref{def:stableEvolution}, any DAQC protocol of this type will be stable. 

Now, we solve the same problem, but by eliminating the last row from the system of equations. This corresponds to the protocol that optimizes the total analog time, whose analog block times are obtained by solving
{\small
\begin{equation}
    \min\,\lVert t\rVert_1, \begin{pmatrix}
        1 & -1 & -1 & 1\\
        1 & -1 & 1 & -1
    \end{pmatrix}\begin{pmatrix}
        t_1\\
        t_2\\
        t_3\\
        t_4
    \end{pmatrix}=T\begin{pmatrix}
        {h_\text{P}}_{12}/{h_\text{P}}_{12}\\
        {h_\text{P}}_{13}/{h_\text{P}}_{13}\\
    \end{pmatrix}, t_i\geq0.
\end{equation}}
By taking the schedule when eliminating the 0's and then adding the characterization errors, we obtain a deviation with in the form of a Hamiltonian $H_\varepsilon$. For our particular example, the Frobenius norm of this Hamiltonian is $\lVert H_\varepsilon\rVert_\text{F}=T\sqrt{2\lVert h_\text{P}\oslash h_\text{S}\rVert_\infty^2+\lVert h_\text{P}\oslash h_\text{S}\rVert_{-\infty}^2}\lVert H_\delta\rVert_\text{F}$. Since the prefactors are again at most linear in the number of qubits, we can prove that DAQC is also stable in this case. However, as expected, the error committed in this case will be larger than the one in which we have applied the error mitigation protocol. This same example is also employed in Apx.~\ref{apx:totalTime}, where we further compare the total analog block time of both protocols.

\section{Discussion about experimental implementations}\label{sec:discussion}

Hardware characterization errors can cause unwanted interactions in any experimental setup. To characterize the  error Hamiltonian, it is necessary to analyze how the different error sources contribute to the hardware characterization error $\delta$. For trapped ions \cite{IonTrapReview}, some error sources to consider are dephasing errors or motional mode heating errors in the trap. They can also experience over-rotation and crosstalk errors induced by laser intensity spillover onto the neighbouring ions, causing a XX-Ising Hamiltonian interactions amongst their neighbouring \cite{Debroy_2020}. Superconducting qubits~\cite{superconductigReview}, suffer from residual coherent couplings between theoretically uncoupled transmons. Another source of errors is the interference of electromagnetic fields that appear between microwave lines. Other sources would be couplings between readout resonators attached to distinct qubits and environmental signal noise that affects all the qubits in a setup~\cite{Sarovar2020detectingcrosstalk}. For Rydberg atoms setups~\cite{RydbergAReview}, characterization error sources can be listed into three main sources~\cite{aquila1}: laser noise, which causes a coherent shot-to-shot variance and time-dependent noise in the Rabi frequency and detuning; atom motion, where atoms experience some thermal motion that changes Rydberg interactions due to the varying distances from their neighbours; and homogeneity caused by imperfect holography of the Rydberg lasers, causing different Rabi frequency and detuning across the 2D array.

The physical systems employed for performing quantum computations often work in time scales where the characterization errors can be interpreted as a constant deviation from the measured Hamiltonian. However, there might be cases in which the effect of noise can cause time dependent calibration errors which are noticeable during the circuit execution. This might be the case, for example, for trapped particles in which vibrations on the particles can change the coupling strength in time scales smaller than the time to run the circuits. In the situation where the system Hamiltonian is affected by high-frequency errors, the errors can be studied the same way as the errors from the couplings outside the system Hamiltonian $(\mathcal{D}\backslash\mathcal{S})$. In this case, the error in the expected value is $\Delta_\mathcal{O}\lesssim 4\,t_\text{A}\,\delta\,\text{deg}(\mathcal{D})\lVert\mathcal{O}\rVert_\text{op}$. However, the error mitigation technique proposed will have no effect on the result as the system Hamiltonian changes at each analog block. As we have previously seen, this dependency still allows us to perform stable estimations of local observables according to Def.~\ref{def:stability} if the error are kept constant with the system size. This allows us to employ DAQC to scale the hardware with a weak constraint on the errors.

\section{Conclusions}\label{sec:conclusions}

Errors in characterizing the coupling strengths in a quantum system lead to errors in the results of a quantum circuit. In the particular case of quantum simulation processes, how the errors grow with the size of the simulator determines its stability. We have shown that quantum simulations implemented with DAQC protocols can be stable under defects in the calibration of the system's interaction Hamiltonian. We additionally calculate the errors in the expectation value of an observable, which is a common task in quantum computing. The obtained results show that, apart from the dependence on the selection of the observable, this error heavily depends on the characteristics of the problem and the system in which we solve it. In particular, there is a direct dependence on the degree and the order of graphs associated to them.

Additionally, we have proposed a new DAQC protocol design for mitigating the calibration errors. This design takes into account the couplings that are assumed to be 0, so that the contribution of possible calibration errors is always suppressed. However, this increased robustness comes at the cost of increasing the total circuit time, and thus, implementing these strategies should be considered on a case-by-case basis. The results obtained in this article extend the possible applicability of DAQC protocols in the near future with softer requirements on the noise scaling. While previously DAQC was envisioned as a practical way of implementing quantum algorithms on NISQ devices, now it opens the door to efficient implementations for larger systems beyond small or intermediate scale setups.

\section*{ACKNOWLEDGMENTS}

MGdA, AA and MS acknowledge financial support from OpenSuperQ+100 (Grant No. 101113946) of the EU Flagship on Quantum Technologies,
also from Spanish Ramón y Cajal Grant No. RYC-2020-030503-I funded by MCIN/AEI/10.13039/501100011033 and by “ERDF A way of making Europe” and “ERDF Invest in your Future”.
This project has also received support from the Spanish Ministry for Digital Transformation and of Civil Service of the Spanish Government through the QUANTUM ENIA project call - Quantum Spain, and by the EU through the Recovery, Transformation and Resilience Plan – NextGenerationEU within the framework of the Digital Spain 2026 Agenda.
We acknowledge funding from Basque Government through Grant No. IT1470-22,
Elkartek project KUBIBIT-kuantikaren berrikuntzarako ibilbide teknologikoak (ELKARTEK25/79),
and the IKUR Strategy under the collaboration agreement between Ikerbasque Foundation and BCAM on behalf of the Department of Education of the Basque Government. 
MGdA acknowledges support from the UPV/EHU and TECNALIA 2021 PIF contract call, 
from the Basque Government through the "Plan complementario de comunicación cúantica" (EXP.2022/01341) (A/20220551), 
and from the Spanish Ministry of Science and Innovation under the Recovery, Transformation and Resilience Plan (CUCO, MIG-20211005). 
AA acknowledges grant PID2021-125823NA-I00 funded by MCIN/AEI/10.13039/501100011033 and by the “European Union NextGenerationEU/PRTR".
AFR acknowledges support from the \"Osterreichischer Wissenschaftsfonds (FWF, Austrian Science Fund) under Grant DOI 10.55776/F71.

\appendix
\section{In depth definition for the graphs associated with the interaction Hamiltonians}\label{apx:definition}

In this work we study two-body Hamiltonians with arbitrary interactions. Instead of working in the Hilbert space of the Hamiltonians $H$, we work in a simpler vector space, in which each Hamiltonian is associated with a vector $h$. For this, we identify uniquely each tuple $(i,j,\mu,\nu)$ with a label in this vector. This vectorization of the Hamiltonians was originally proposed in \cite{Adrian2020DAQC}, and in a general form in \cite{MikelAlvaro2024}.

Now, we go further, and associate a graph with a given two-body Hamiltonian. In particular, we define a weighted multigraph for describing the connectivity of each of the Hamiltonians, $H\leadsto\mathcal{H}$. A general Hamiltonian can be written as
\begin{equation}
    H=\sum_{\{\{i,j\},\{\mu,\nu\}\}\in \mathcal{D}}h_{ij}^{\mu \nu}\,\sigma^\mu_i\sigma^\nu_j.
\end{equation}
The vertices of $\mathcal{H}$ correspond to each of the qubits, $V_\mathcal{H}=\{i\}_{i=1}^N$. The edge for a given coupling $h_{ij}^{\mu\nu}$ is defined between the corresponding vertices $(i,j)$ with $i<j$, and are labeled with the interaction axis $\mu\nu\in\{x,y,z\}$. As a particular example, the most general Hamiltonian composed of ATA interactions in all possible axes for $N$ qubits have an associated complete multigraph of order 9, this is $9K_N$.  

These definitions can be directly applied to both the system Hamiltonian $(H_\text{S}\leadsto\mathcal{S})$ and the problem Hamiltonians $(H_\text{P}\leadsto\mathcal{P})$. We divide the Hamiltonian for the measurement defects into two, one with significant defects ($H_\delta\leadsto\mathcal{D}$)
\begin{equation}\label{eq:defectHam}
    H_\delta=\sum_{\{\{i,j\},\{\mu,\nu\}\}\in \mathcal{D}}{h_\delta}_{ij}^{\mu \nu}\,\sigma^\mu_i\sigma^\nu_j,
\end{equation}
with $\lvert{h_\delta}_{ij}^{\mu \nu}\rvert\leq\delta$, and the other with negligible defects ($H_\eta\leadsto\mathcal{M}$)
\begin{equation}
    H_\eta=\sum_{\{\{i,j\},\{\mu,\nu\}\}\in \mathcal{M}}{h_\eta}_{ij}^{\mu \nu}\,\sigma^\mu_i\sigma^\nu_j,
\end{equation}
with $\lvert{h_\eta}_{ij}^{\eta \nu}\rvert\leq\mu$ such that $\eta\ll\delta$. The condition on the coupling strength can be also written as, $\lVert h_\delta\rVert_\infty=\delta$. The two graphs associated with the defect Hamiltonians are complementary in the sense that the union of both forms the complete graph of $N$ nodes, $\mathcal{D}\cup\mathcal{M}=K_N$. The relationship between the rest of the graphs is $\mathcal{P}\subseteq\mathcal{S}\subseteq\mathcal{D}$, in the sense that all graphs have the same set of vertices, $V_\text{P}=V_\text{S}=V_\delta$, but the set of edges indeed have this relationship $E_\text{P}\subseteq E_\text{S}\subseteq E_\delta$. This property assures that there is no coupling in the problem Hamiltonian without a respective non-zero coupling in the source Hamiltonian, ${h_\text{S}}_{ij}^{\mu \nu}=0\Rightarrow{h_\text{P}}_{ij}^{\mu \nu}=0$. Since it will be useful later, we define now the degree of these graphs as $\deg(\mathcal{P})\leq\deg(\mathcal{S})\leq\deg(\mathcal{D})\leq N-1$ and $\deg(\mathcal{M})<N-1$. Note that $\mathcal{P}$, $\mathcal{S}$ and $\mathcal{D}$ are connected graphs, while $\mathcal{M}$ may not necessarily be connected.

\begin{figure}
    \centering
    \includegraphics[width=0.4\linewidth]{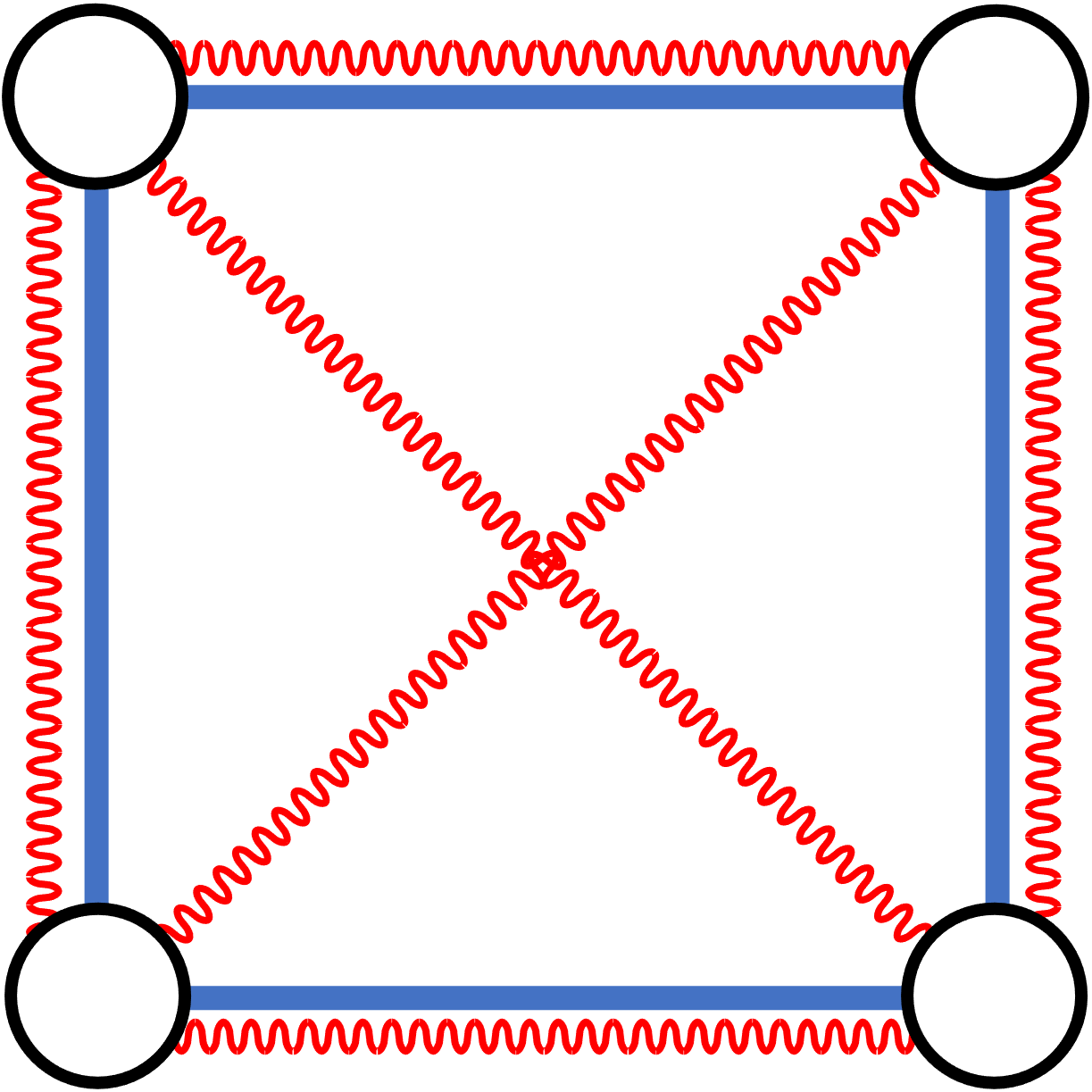}
    \caption{Representation of a hardware with a square connectivity $\mathcal{S}$ (blue) and all-to-all characterization errors $\mathcal{D}$ (red). Taking this as the unit cell of a square lattice, we can represent the second example given in Sec.~\ref{sec:stability}. For this example, we can pictorially check that $\mathcal{S}\subset\mathcal{D}$ and $\text{deg}(\mathcal{S})<\text{deg}(\mathcal{D})$.}
    \label{fig:enter-label}
\end{figure}

\section{Bound for the p-norm of the vector of couplings}\label{apx:pnormvectorcouplings}

We start from a usual DAQC schedule to simulate a target Hamiltonian $H_\text{P}$ with a source Hamiltonian $H_\text{S}$. We vectorize the coupling constants for each Hamiltonian as usual, $h_\text{P}$ and $h_\text{S}$ respectively. Then, the schedule for simulating $H_\text{P}$ for a time $T$ is obtained from
\begin{equation}\label{eqApx:scheduleIdealCompacto}
    M\big\rvert_\mathcal{S}\,t=T\,(h_\text{P}\oslash h_\text{S})\big\rvert_\mathcal{S},
\end{equation}
where $\oslash$ denotes the Hadamard division or elementwise division. We also employ the Hadamard product or elementwise product throughout the appendix \footnote{For $a=\{a_i\}$ and $b=\{b_i\}$, $a\odot b=c=\{c_i\}=\{a_i\cdot b_i\}$.}. Here, we have employed the usual way of dealing with indeterminate forms, this is, the couplings that do not appear in the source Hamiltonian are taken out of the equation. We denote that we are working in this restricted subspace by using $\cdot\big\rvert_\mathcal{S}$.

Now, we assume that the real hardware $H_\text{S}'$ has some deviation from the initial values $H_\delta$, such that  $H_\text{S}'=H_\text{S}+H_\delta$ or $h_\text{S}'=h_\text{S}+h_\delta$. Note here that $h_\delta$ has support on $\mathcal{D}$, so we can differentiate two subspaces, one on $\mathcal{S}$ and the other on $\mathcal{D}\backslash\mathcal{S}$. This way we have $h_\delta=h_\delta\big\rvert_\mathcal{S}+h_\delta\big\rvert_{\mathcal{D}\backslash\mathcal{S}}$ and equivalently $h_\varepsilon=h_\varepsilon\big\rvert_\mathcal{S}+h_\varepsilon\big\rvert_{\mathcal{D}\backslash\mathcal{S}}$.
The simulated Hamiltonian will deviate from the expected Hamiltonian by some error, $H_\text{P}'=H_\text{P}+H_\varepsilon$ or $h_\text{P}'=h_\text{P}+h_\varepsilon$. 

If we restrict ourselves to $\mathcal{S}$, the schedule we are employing is identical to the previous one used in Eq.~\ref{eqApx:scheduleIdealCompacto}, so we can write
\begin{equation}\label{eq:DAQC_vecS}
    M\big\rvert_\mathcal{S}\,t=T\, (h_\text{P}\oslash h_\text{S})\big\rvert_\mathcal{S}=T((h_\text{P}+h_\varepsilon)\oslash(h_\text{S}+h_\delta))\big\rvert_\mathcal{S}.
\end{equation}
Using the distributivity of the Hadamard division and simplifying the simulation time $T$, we can write
\begin{equation}
    \left(h_\text{P}\oslash h_\text{S}-h_\text{P}\oslash(h_\text{S}+h_\delta)\right)\big\rvert_\mathcal{S}=\left(h_\varepsilon\oslash(h_\text{S}+h_\delta)\right)\big\rvert_\mathcal{S}.
\end{equation}
Since we are working with elementwise divisions, we can treat these expressions as numbers, and thus, we can simplify the expression using the Hadamard product or elementwise product as
\begin{equation}
    \left(h_\text{P}\odot(h_\text{S}+h_\delta)\oslash h_\text{S}-h_\text{P}\right)\big\rvert_\mathcal{S}=h_\varepsilon\big\rvert_\mathcal{S}.
\end{equation}
By further simplifying, we obtain a compact expression for the vector of couplings for the error Hamiltonian,
\begin{equation}\label{eq:heSwithoutError}
    h_\varepsilon\big\rvert_\mathcal{S}=h_\text{P}\odot h_\delta\oslash h_\text{S}\big\rvert_\mathcal{S}.
\end{equation}
We can find a bound for the $p$-norm of this vector, for which we assume that the maximum error in measuring each coupling is $\delta$, $\lVert h_\delta\rVert_\infty=\delta$. The maximum error in the simulated Hamiltonian will be reached when the measurement error is maximal,
\begin{equation}\label{eq:heS}
    \lVert h_\varepsilon\rVert_p\big\rvert_{\mathcal{S}}=\lVert h_\text{P}\odot h_\delta\oslash h_\text{S}\rVert_p\big\rvert_{\mathcal{S}}\leq\delta\lVert h_\text{P}\oslash h_\text{S}\rVert_p\big\rvert_{\mathcal{S}}.
\end{equation}

The expression for the couplings in $\mathcal{D}\backslash\mathcal{S}$ is calculated just with the error Hamiltonians
\begin{equation}\label{eq:DAQC_vecDS}
    M\big\rvert_{\mathcal{D}\backslash\mathcal{S}}\,t=T(h_\varepsilon\oslash h_\delta)\big\rvert_{\mathcal{D}\backslash\mathcal{S}}.
\end{equation}
From this expression we can obtain the error
\begin{equation}\label{eq:heDS}
    h_\varepsilon\big\rvert_{\mathcal{D}\backslash\mathcal{S}}= \frac{1}{T} M t\odot h_\delta\big\rvert_{\mathcal{D}\backslash\mathcal{S}}.
\end{equation}

Now we calculate a bound for the $p$-norm of this vector. In the worst case, the contribution of each term is maximal, and so it depends on the number of edges of the graph corresponding to the defect Hamiltonian in the couplings belonging with $\mathcal{D}\backslash\mathcal{S}$
\begin{equation}\label{eq:errorDs}
    \lVert h_\varepsilon\rVert_p\big\rvert_{\mathcal{D}\backslash\mathcal{S}}=\lVert M t \odot h_\delta/T\rVert_p\big\rvert_{\mathcal{D}\backslash\mathcal{S}}\leq\delta \dfrac{t_\text{A}}{T} \lvert E_{\mathcal{D}\backslash\mathcal{S}}\rvert^{1/p}.
\end{equation}

By combining Eq.~\ref{eq:heS} and Eq.~\ref{eq:errorDs}, we arrive at the result
\begin{equation}
    \lVert h_\varepsilon\rVert_p \leq\delta\lVert h_\text{P}\oslash h_\text{S}\rVert_p\big\rvert_\mathcal{S}+\delta \dfrac{t_\text{A}}{T} \lvert E_{\mathcal{D}\backslash\mathcal{S}}\rvert^{1/p},
\end{equation}
where the first term is only evaluated within the graph $\mathcal{S}$ and $\lvert E_{\mathcal{D}\backslash\mathcal{S}}\rvert$ is the number of edges in the graph difference of $\mathcal{D}$ and $\mathcal{S}$ defined as the graph obtained by removing the edges of $\mathcal{S}$ from $\mathcal{D}$.

\section{Bound for the operator norm for the error Hamiltonian}\label{apx:operatornorm}

We want to obtain a bound for the error Hamiltonian $H_\varepsilon=H'_\text{P}-H_\text{P}$ in terms of the operator norm. We start by writing $H_\varepsilon$ as a sum of Hamiltonian terms,
\begin{equation}
    \lVert H_\varepsilon\rVert_\text{op}=\lVert \sum_{\{\{i,j\},\{\mu,\nu\}\}\in \mathcal{D}}{h_\varepsilon}_{ij}^{\mu \nu}\,\sigma^\mu_i\sigma^\nu_j\rVert_\text{op}.
\end{equation}
Using the triangular inequality, we have an upper bound
\begin{equation}
    \lVert H_\varepsilon\rVert_\text{op}\leq \sum_{\{\{i,j\},\{\mu,\nu\}\}\in \mathcal{D}}\lvert {h_\varepsilon}_{ij}^{\mu \nu}\rvert\lVert\sigma^\mu_i\sigma^\nu_j\rVert_\text{op}.
\end{equation}
The operator norm of the two-body interactions is identically equal to 1 for all combinations of pairs of qubits. Then, we can identify the resulting expression with the 1-norm of the vector of couplings of the error Hamiltonian,
\begin{equation}\label{eqAPX:HamNormOpFirst}
    \lVert H_\varepsilon\rVert_\text{op}\leq\lVert h_{\varepsilon}\rVert_1.
\end{equation}
Finally, particularizing the bound from Apx.~\ref{apx:pnormvectorcouplings} for the 1-norm, we have 
\begin{equation}
     \lVert H_\varepsilon\rVert_\text{op}\leq\delta\lVert h_\text{P}\oslash h_\text{S}\rVert_1\big\rvert_\mathcal{S}+\delta \dfrac{t_\text{A}}{T} \lvert E_{\mathcal{D}\backslash\mathcal{S}}\rvert.
\end{equation}
We can use the lower bound for the operator norm of the calibration error two-body Hamiltonian, $\lVert H_\delta\rVert_\text{op}\geq \delta$, to relate this equation with the stability,
\begin{equation}\label{eqAPX:errorHamiltonianOpNorm}
     \lVert H_\varepsilon\rVert_\text{op}\leq\left(\lVert h_\text{P}\oslash h_\text{S}\rVert_1\big\rvert_\mathcal{S}+\dfrac{t_\text{A}}{T} \lvert E_{\mathcal{D}\backslash\mathcal{S}}\rvert\right)\lVert H_\delta\rVert_\text{op}.
\end{equation}

\begin{figure}[t]
    \centering
    \includegraphics[width=\linewidth]{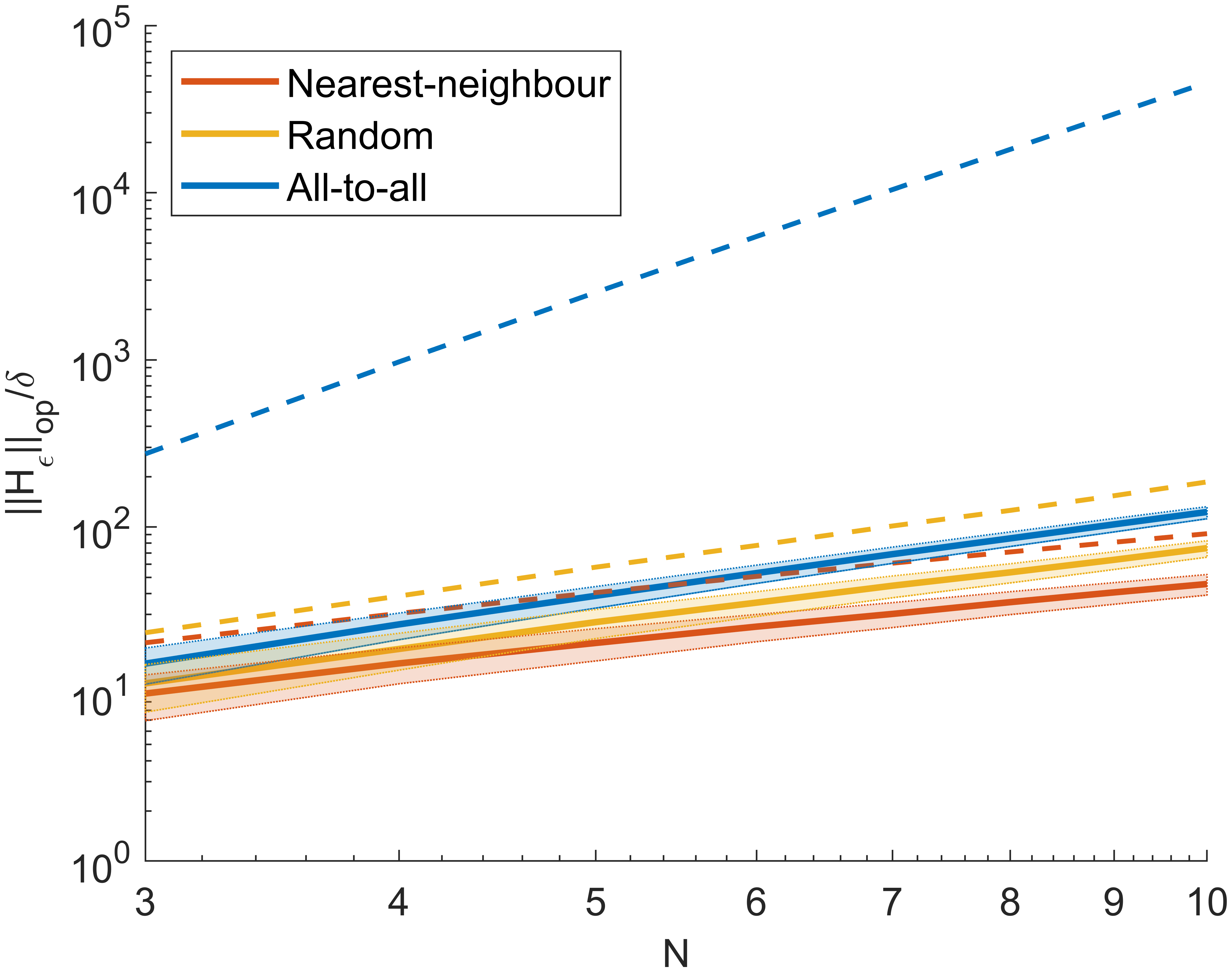}
    \caption{Repetition of the simulations for Fig.~\ref{fig:bound} but for the characterization error mitigated DAQC protocol. As the second term in Eq.~\ref{eq:errorHamiltonian} is taken out, the bounds are tighter compared to the simulation results.}
    \label{fig:boundmitigated}
\end{figure}

For verifying these bounds, we have run a series of simulations. The simulation results shown in Fig.~\ref{fig:bound} are calculated from 10000 different problems for each data point. In these simulations we have generated problems with different topologies. For simplicity, we have assumed that the topology of the problem and source Hamiltonians are the same, $\mathcal{S}=\mathcal{P}$. For the nearest-neighbour (NN) problems we have included errors up to the second neighbours. For the random problems, we first have generated a NN problem. Then, each remaining edge has a probability of 0.2 to be included. In all cases, the characterization error Hamiltonian is all-to-all (ATA). For the ATA problems, all edges from the complete graph appear in all graphs. The evolution time is fixed to $T=1$ in arbitrary units. For the generation of the coupling constants for the Hamiltonian, we uniformly sample from a set range of $[g/2,3g/2]$ and we pick the sign by throwing a 50/50 coin. For the characterization error couplings we uniformly sample from the range $[-\delta,\delta]$, for a value of $\delta=10$. In all figures shown in this paper, we show the results for a value of $g=100$ in arbitrary units for both the problem and the source Hamiltonians. Although not shown here, similar results can be obtained by modifying the proportion between the $g$ value of the problem and the source Hamiltonians. In the main text, Fig.~\ref{fig:bound} shows the simulations for the case in which we employ the previous DAQC protocol for dealing with $0/0$ indefinite forms in $h_\text{P}\oslash h_\text{S}$. On the other hand, Fig.~\ref{fig:boundmitigated} shows the results for the case in which the error mitigation protocol is employed. Comparing both we can see that the bounds for the error mitigated case are tighter. Additionally, the addition of the error mitigation technique reduces the operator norm of the error Hamiltonian.

\section{Bound for the Frobenius norm for the error Hamiltonian}\label{apx:frobeniusnorm}

Similarly as in the previous appendix, we want to bound the Frobenius norm. For this, we will first write the full expression for the error Hamiltonian
\begin{equation}
    \lVert H_\varepsilon\rVert_\text{F}=\lVert \sum_{\alpha\in\mathcal{D}}\sum_k\frac{t_k}{T}M_{\alpha,k}{h_\delta}_{\alpha}\sigma_i^\mu\sigma_j^\nu\rVert_\text{F},
\end{equation}
where we have directly use $\alpha$ to label the couplings. Here, we identify all the terms summing on the index $k$ with $\sum_k\frac{t_k}{T}M_{\alpha,k}= c_\alpha$, and we use the definition of the Frobenius norm. For simplicity in the notation, we will work with the square of the Frobenius norm
\begin{equation}
    \lVert H_\varepsilon\rVert_\text{F}^2=\text{Tr}\left( \sum_{\{\alpha,\beta\}\in\mathcal{D}}c_\alpha c_\beta {h_\delta}_{\alpha}{h_\delta}_{\beta}\sigma_i^\mu\sigma_j^\nu\sigma_{i'}^{\mu'}\sigma_{j'}^{\nu'}\right).
\end{equation}
Using the linearity of the trace we rewrite this expression by taking all the constants out of the trace
\begin{equation}\label{eq:Hsinclussion}
    \lVert H_\varepsilon\rVert_\text{F}^2=\sum_{\{\alpha,\beta\}\in\mathcal{D}}c_\alpha c_\beta {h_\delta}_{\alpha}{h_\delta}_{\beta} \text{Tr}\left( \sigma_i^\mu\sigma_j^\nu\sigma_{i'}^{\mu'}\sigma_{j'}^{\nu'}\right).
\end{equation}
By noting that the operator here will be the identity only if $\alpha=\beta$ and zero otherwise, we obtain
\begin{equation}
    \lVert H_\varepsilon\rVert_\text{F}^2=\sum_{\{\alpha,\beta\}\in\mathcal{D}}c_\alpha c_\beta {h_\delta}_{\alpha}{h_\delta}_{\beta} 2^N\delta_{\alpha\beta}=\sum_{\alpha\in\mathcal{D}}c_\alpha^2 {h_\delta}_\alpha^2 2^N,
\end{equation}
where here $\delta_{\alpha\beta}$ is the Kronecker delta function. As done in Apx.~\ref{apx:pnormvectorcouplings}, we can identify part of this sum with the system of equations employed to obtain DAQC protocol, 
\begin{equation}
    \lVert H_\varepsilon\rVert_\text{F}^2 =\sum_{\alpha\in\mathcal{S}}\left(\frac{{h_\text{P}}_\alpha}{{h_\text{S}}_\alpha}\right)^2{h_\delta}^2 2^N +
    \sum_{\alpha\in\mathcal{D}/\mathcal{S}}c_\alpha^2 {h_\delta}_\alpha^2 2^N.
\end{equation}
As before, since we do not have information about the sign of the couplings outside the source Hamiltonian, we will bound the second term with the total analog block time $t_\text{A}$. We will also use the bound for the defect Hamiltonian $\lVert h_\delta\rVert_\infty= \delta$,
\begin{equation}
    \lVert H_\varepsilon\rVert_\text{F}^2 \leq 
    \lVert h_\text{P} \oslash h_\text{S} \rVert_2^2\big\rvert_\mathcal{S} \delta^2 2^N +
    \lvert E_{\mathcal{D}/\mathcal{S}}\rvert \left(\frac{t_\text{A}}{T}\right)^2 \delta^2 2^N.
\end{equation}
Here we can identify the Frobenius norm of the defect Hamiltonian, and write
\begin{equation}
    \lVert H_\varepsilon\rVert_\text{F}^2\leq \left(\lVert h_\text{P} \oslash h_\text{S} \rVert_2^2\big\rvert_\mathcal{S} +
    \lvert E_{\mathcal{D}/\mathcal{S}}\rvert \left(\frac{t_\text{A}}{T}\right)^2 \right)\lVert H_\delta\rVert_\text{F}^2.
\end{equation}
This relation for 2-body Hamiltonians comes from the fact that we can find a lower bound of the Frobenius norm for a 2-body Hamiltonian,
\begin{equation}
    \lVert H_\delta\rVert_\text{F}\geq 2^{N/2}\lVert H_{\delta,\text{min}}\rVert_\text{op}\geq  2^{N/2} \delta,
\end{equation}
which corresponds to a Hamiltonian with a single term with coupling $\delta$, such that the Hamiltonian still fulfills the assumption that $\lVert h_\delta\rVert_\infty=\delta$.

\section{Bound for the error in the expectation value}\label{apx:stability}

In this appendix, we calculate the error in the expectation value of an observable when using DAQC in the presence of calibration defects, as defined in Def.~\ref{def:stability}.

\subsection{Error in the expectation value through the commutator $[\mathcal{O},H_\text{P}]$}

To bound this expression, we turn to the following bound of the observables' expected values for an arbitrary two-body Hamiltonian from \cite{adrian2023},

\begin{equation}
\begin{split}
\left\|e^{i H_\text{P} T} \mathcal{O} e^{-i H_\text{P} T}-e^{i H_\text{P}^{\prime} T} \mathcal{O} e^{-i H_\text{P}^{\prime} T}\right\|_\text{op} \\
\leq \begin{cases}
T\left\|\left[H_\varepsilon, \mathcal{O}\right]\right\|_\text{op} &\text {if }\left[H_\text{P}, H_\text{P}^{\prime}\right]=0,  \\
2 T\|\mathcal{O}\|_\text{op}\left\|H_\varepsilon\right\|_\text{op} &\text {else},
\end{cases}
\end{split}
\end{equation}
where we simply need to use the trace's cyclic property to reorganize the equation into
\begin{equation}
\begin{split}
    \Delta_\mathcal{O}=\left\lvert \text{Tr}( \rho_f (e^{i H_\text{P} T} \mathcal{O} e^{-i H_\text{P} T}-e^{i H_\text{P}^{\prime} T} \mathcal{O} e^{-i H_\text{P}^{\prime} T}))\right\rvert \\
    \leq \begin{cases}
    T\left\|\left[H_{\varepsilon}, \mathcal{O}\right]\right\|_\text{op}=O(T \delta \operatorname{deg} \mathcal{D}) &\text {if }\left[H_\text{P}, H_\text{P}^{\prime}\right]=0,  \\
    2 T\|\mathcal{O}\|_\text{op}\left\|H_{\varepsilon}\right\|_\text{op} &\text {else}.
\end{cases}
\end{split}
\end{equation}
For the first case, the expression depends only on the simulation time, the error and the degree of $\mathcal{D}$, $\Delta_\mathcal{O}=O(T\delta \text{deg}\mathcal{D})$. In this case, the expression is independent of the system size given a local Hamiltonian $H_\delta$. However, in the second case, for noncommuting $H_P, H'_P$, the bound for $\Delta_\mathcal{O}$ depends on the operator norm of the error Hamiltonian (Eq.~\ref{eqAPX:errorHamiltonianOpNorm}). This can be problematic, since in general we can only upper bound this expression, via the $\infty$-norm, by a system size-dependent quantity, $\Delta_\mathcal{O}\leq 2T\delta N\lVert\mathcal{O}\rVert_\text{op}\lVert h_\text{P}\oslash h_\text{S}\rVert_\infty$. If we however assume that the Hamiltonians are geometrically local, the argument of \cite{adrian2023}, based on Lieb-Robinson bounds, is enough to recover the stability of DAQC even in the non-commuting case, provided that we are in a regime where Eq.~\eqref{eq:approxDAQC} holds (which may require scaling $q$ with the system size).

\subsection{Error in the expectation value in the small defect limit}

To recover a useful expression for the non-commuting case, we introduce an extra assumption. We calculate the error in the expectation value assuming $\delta\ll\left\lVert h_\text{S}\right\rVert_{-\infty}$, this is, we assume that the calibration error is much smaller than the couplings in $H_\text{S}$ \footnote{This bound is obtained by ordering the implementation of the digital-analog blocks from largest to the lowest block times. However, we can find a less restrictive bound if we do not impose such condition, $\delta\ll\lVert h_\text{P}\rVert_\infty\lVert h_\text{S}\rVert_\infty/\lVert h_\text{P}\rVert_{-\infty}$. However, the latter bound is mathematically motivated while the one in the main text is more physically motivated. The conclusions do not depend of the choice of these bounds.}. By using the Trotter formula, we can divide the real evolution into two terms $U'\approx U_\delta U$, where $U=\prod_k U_k$ is the ideal evolution designed with the measured parameters of $H_\text{S}$ and $U_\delta=\prod_k e^{-i t_k H_\delta^{(k)}}\approx e^{-iTH_\varepsilon}$ is the extra evolution generated by the defect Hamiltonians. The error of this decomposition calculated as operator norm of the difference between both expressions is in $\mathcal{O}(T^2\lVert H_\delta\rVert_\text{op}\lVert H_\text{S}\rVert_\text{op})$. This error is negligible compared to $\Delta_\mathcal{O}$ because we are already working in the small defect limit, $\lVert H_\delta\rVert_\text{op}\ll\lVert H_\text{S}\rVert_\text{op}$, if we additionally assume we are working in a short time regime that fulfills $T\lVert H_\text{S}\rVert_\text{op}\ll 1$, which also implies that Eq.~\ref{eq:approxDAQC} holds.

Now, we repeat the calculation by following lemma 2 proof in~\cite{adrian2023}. We identify $U U'^\dagger\mathcal{O}{U'} U^\dagger\approx U_\delta\mathcal{O}U_\delta^\dagger\approx e^{-iTH_\varepsilon}\mathcal{O}e^{itH_\varepsilon}$. Therefore, when computing the derivative of this quantity to respect to $T$, we obtain the following expression
\begin{equation}
    \Delta_\mathcal{O}\leq T\left\|[H_\varepsilon,\mathcal{O}]\right\|_\text{op}.
\end{equation}
Let us start by considering a single-body observable on the first qubit. Using Eqs.~\ref{eq:heSwithoutError}, \ref{eq:heDS} we expand the commutator,
\begin{equation}
\begin{split}
    \Delta_\mathcal{O}&\leq T\rVert[\sum_{\beta\in\mathcal{S}}\frac{h_{P,\beta}}{h_{S,\beta}}\delta_\beta\,\sigma_i^\mu\sigma_j^\nu,\mathcal{O}]\\
    &\phantom{=}+[\sum_{\gamma\in\mathcal{D}\backslash\mathcal{S},k}t_k\,M_{\gamma,k}\,\delta_{\gamma}\,\sigma_\ell^\mu\sigma_m^\nu,\mathcal{O}]\rVert_\text{op}
\end{split}
\end{equation}
For simplicity, let us use the linearity of the trace to study the first term first (${\Delta_\mathcal{O}}_1$). In this term, we see that the commutator is zero for the cases in which the couplings and the observable have different supports,
\begin{equation}
    {\Delta_\mathcal{O}}_1=\lVert[\sum_{\beta\in\mathcal{S'}}\frac{T\, h_{P,\beta}}{h_{S,\beta}}\delta_\beta\,\sigma_i^\mu\sigma_j^\nu,\mathcal{O}]\rVert_\text{op},
\end{equation}
where $\mathcal{S}'$ only have the edges with support on $\text{supp}(\mathcal{O})$. Now, let us compute the commutator and employ the triangular inequality twice
\begin{equation}
\begin{split}
    {\Delta_\mathcal{O}}_1&\leq2\rVert\sum_{\beta\in\mathcal{S}'} \frac{T h_{\text{P},\beta}}{h_{\text{S},\beta}}\delta_\beta a_\eta \sigma_i^\mu\sigma_j^\nu\mathcal{O}\lVert_\text{op}\\
    &\leq 2\sum_{\beta\in\mathcal{S}'} \left|\frac{T h_{\text{P},\beta}}{h_{\text{S},\beta}}\delta_\beta a_\eta \right|\lVert\sigma_i^\mu\sigma_j^\nu\mathcal{O}\lVert_\text{op}
\end{split}
\end{equation}
Now, we can equate the operator norm of the Pauli operators by 1, $\lVert h_\delta\rVert_\infty=\delta$, and use the fact that the operator norm is sub-multiplicative \footnote{For sub-multiplicative norms $\lVert AB\rVert\leq\lVert A\rVert\,\lVert B\rVert$},
\begin{equation}
    {\Delta_\mathcal{O}}_1\leq 6\sum_{\beta\in\mathcal{S}'} \left|\frac{T h_{\text{P},\beta}}{h_{\text{S},\beta}}\delta \right|=6T\,\delta\lVert\mathcal{O}\rVert_\text{op}\lVert h_\text{P}\oslash h_\text{S}\rVert_i\big\rvert_{\mathcal{S}'}.
\end{equation}
Here, the 1-norm of the vectorized Hamiltonian is only taken with the elements within the subgraph $\mathcal{S}'$, which has a maximum number of elements of $\text{supp}(\mathcal{O})\text{deg}(\mathcal{S})$. With this, we arrive at this first term for the expression
\begin{equation}
    {\Delta_\mathcal{O}}_1\leq 6T\,\delta\,\text{supp}(\mathcal{O})\text{deg}(\mathcal{P})\lVert\mathcal{O}\rVert_\text{op}\lVert h_\text{P}\oslash h_\text{S}\rVert_\infty,
\end{equation}
which is similar to the previous expression but where now we have to take into account the degree of the multigraph of the non-zero interactions of $H_\text{P}$. 

Let us compute now the second term. By following a similar calculation, we can bound obtain a bound by counting the number of couplings in $\mathcal{D}\backslash\mathcal{S}$ connected to the measured qubits. Thus, the error on the expectation value of an observable is given by
\begin{multline}
    \Delta_\mathcal{O}\leq\,6T\delta\,\text{supp}(\mathcal{O})\text{deg}(\mathcal{P})\lVert\mathcal{O}\rVert_\text{op}\lVert h_\text{P}\oslash h_\text{S}\rVert_\infty\\
    +6t_\text{A}\delta\,\text{supp}(\mathcal{O})\text{deg}(\mathcal{D}\backslash\mathcal{S})\lVert\mathcal{O}\rVert_\text{op}.
\end{multline}
{Here, the terms that could be dependent on the system size are $\text{supp}(\mathcal{O})$, $\lVert\mathcal{O}\rVert_\text{op}$, $\text{deg}(\mathcal{D}\backslash\mathcal{S})$ and $t_\text{A}$. For the first two, we can impose that $\mathcal{O}$ is a $k$-local operator, and for the third, we can force the associated graph to have bounded degree. For the latter, it was conjectured that $t_\text{A}$ could depend at most linearly with the system size \cite{Bassler2024TimeOptimal}.}

\section{Total analog block times changes with the way of dealing with indeterminate forms}\label{apx:totalTime}

In some DAQC problems, both the problem and the source Hamiltonians are identically 0, so an indeterminate form of the type $0/0$ appears in the problem of obtaining the analog block times. 

Previously, this was solved by taking out the corresponding row from the problem vector $b$ and from the matrix $M$. Then, for this case, the problem of finding the optimal DAQC schedule is
\begin{equation}
    \min \lVert t\rVert_1 \text{ s.t. }Mt=b, t\geq0,
\end{equation}
for $b\in\mathbb{R}^N$ and $M\in\{\pm1\}^{N\times K}$. For solving the minimization problem, one explores a space with a number of parameters equal to the number of free parameters for the solution of the system of equations, in this case, $K-N$. Assume that we find a solution that gives a global minimum for this problem, $t_\text{opt}$. Note that we do not require the solution to be unique. 

Let us study the case in which we leave the $0/0$ indeterminate forms of the problem as arbitrary parameters. We reorder the couplings such that the indeterminate forms goes into the last indices. For this case, the problem of finding a DAQC schedule is 
\begin{equation}
    \min_t \lVert t\rVert_1 \text{ s.t. } \left(\begin{array}{c|c}
        M & A\\ \hline
        B & C
        \end{array}\right)t =\begin{pmatrix}
        b\\
        k
    \end{pmatrix},
\end{equation}
where $b$ is the vector with the division of $h_\text{P}\oslash h_\text{S}$ which gives a real number, and $k$ is a vector with arbitrary real numbers and are associated with the couplings where an indeterminate form appeared. The submatrix $A\in\{\pm1\}^{N\times y}$ can contain columns that also appear in the submatrix $M$. The number of rows in $B\in\{\pm1\}^{x\times K}$ and $C\in\{\pm1\}^{x\times y}$ is the same as the number of elements in $k$, $\lvert k\rvert=x$. 

Now, we solve a slightly different problem, in which we use the free parameters in $k$ to further optimize the schedule,
\begin{equation}
    \min_{t,k} \lVert t\rVert_1 \text{ s.t. } \left(\begin{array}{c|c}
        M & A\\ \hline
        B & C
        \end{array}\right)t =\begin{pmatrix}
        b\\
        k
    \end{pmatrix}, t\geq0.
\end{equation}
The optimal total time corresponds to a situation in which no extra time is added with respect to the original problem. This is the same as the problem of finding $k$ for the following system of equations
\begin{equation}
    \left(\begin{array}{c|c}
        M & A\\ \hline
        B & C
    \end{array}\right) 
    \begin{pmatrix}
            t_\text{opt}\\
            0
    \end{pmatrix} =
    \begin{pmatrix}
        b\\
        k
    \end{pmatrix} 
    \Rightarrow B t_\text{opt}=k.
\end{equation}
The vector $k$ is uniquely defined by $B$ and $t_\text{opt}$, and even though the optimal solution $t_\text{opt}$ may not be unique, with high probability the elements of $k$ will be distinct from 0.

Thus, when we want to calculate the schedule incorporating the calibration error mitigation technique, this is, when solving the problem for all indeterminate forms equal to 0, the optimization problem is
\begin{equation}
    \min_{t} \lVert t\rVert_1 \text{ s.t. } \left(\begin{array}{c|c}
        M & A\\ \hline
        B & C
        \end{array}\right)t =\begin{pmatrix}
        b\\
        0
    \end{pmatrix}, t\geq0.
\end{equation}
Then, the total time for this situation is larger or equal to that in the situation where we remove the indeterminate forms from the equations,
\begin{equation}
    t_{A,\text{remove}}\leq t_{A,\text{0's}}.
\end{equation}

\textit{Example with 3 qubits:} Let us employ the same example as the one of Sec.~\ref{sec:mitigation}, but now focusing on the total annealing time. Assume we have a system with 3 qubits but only two couplings. If we solve the problem by assigning 0 to the missing coupling, we have to solve the following problem
\begin{equation}
    \min \sum_{i=1}^4 t_i, \begin{pmatrix}
        1 & -1 & -1 & 1\\
        1 & -1 & 1 & -1\\
        1 & 1 & -1 & -1
    \end{pmatrix}\begin{pmatrix}
        t_1\\
        t_2\\
        t_3\\
        t_4
    \end{pmatrix}=\begin{pmatrix}
        b_1\\
        b_2\\
        0
    \end{pmatrix}, t_i\geq0.
\end{equation}
In this case, the solution to the problem yields a total time of $t_{A,\text{put 0's}}=\lvert b_1\rvert+\lvert b_2\rvert=\lVert b\rVert_1$.

Now, we solve the same problem, but by eliminating the last row. The remaining columns of $M$ now have all $4$ combinations of $\{\pm1\}^2$. Solving the problem now yields a solution with $t_{A,\text{remove}}=\max(\lvert b_1\rvert,\lvert b_2\rvert)=\lVert b\rVert_\infty$.

Thus, in this trivial case, we see that the analog block time when substituting the indeterminate forms with 0 is longer or equal to the times when we eliminate the corresponding row, $t_{A,\text{remove}}\leq t_{A,\text{put 0's}}$. 

\begin{figure}[t]
    \centering
    \includegraphics[width=\linewidth]{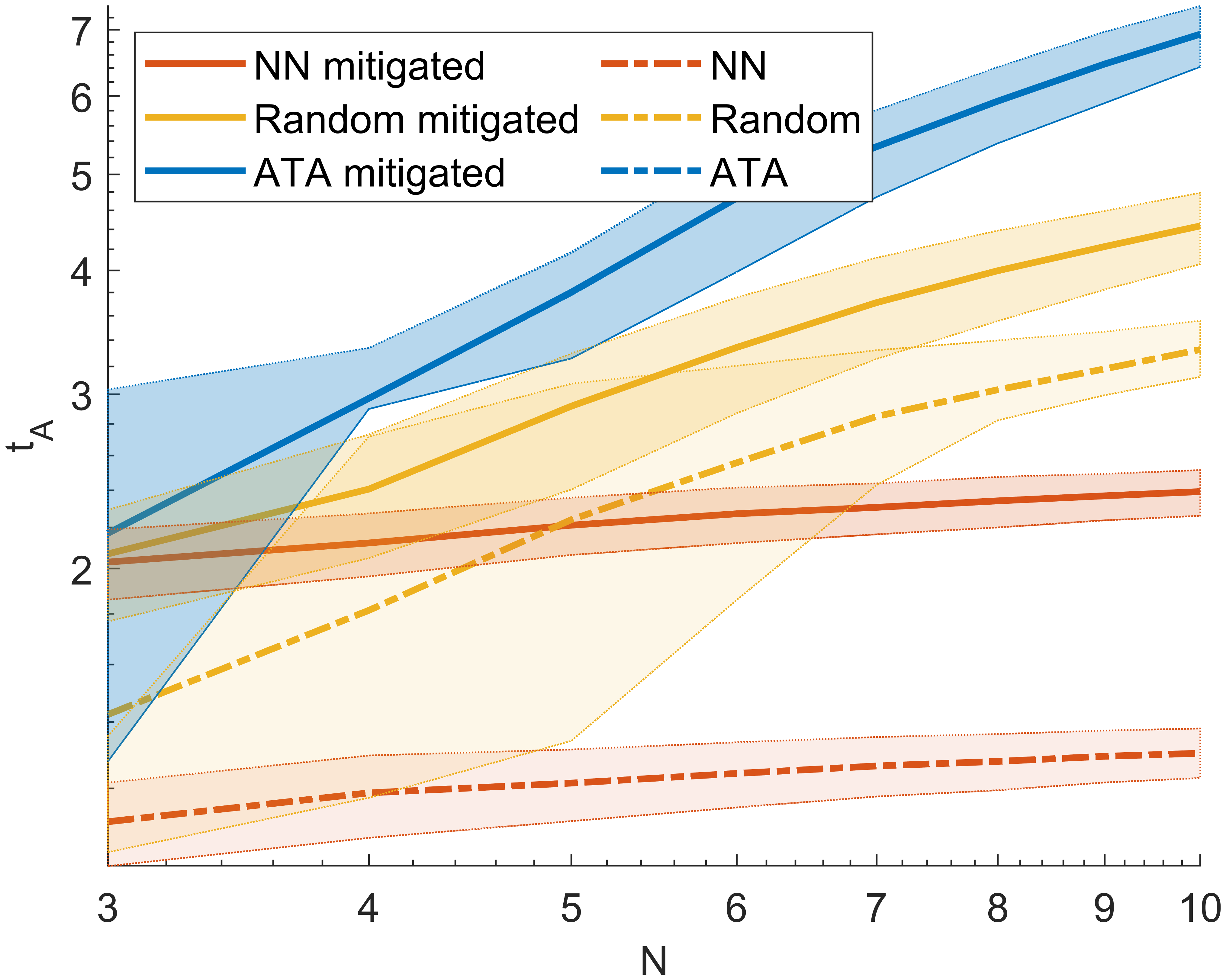}
    \caption{Total analog time $t_\text{A}$ for a Hamiltonian simulation task with the 3 different problem topologies. We show the times for both the usual DAQC protocols and the one with the error mitigation technique. The results for the ATA and the ATA with error mitigation overlap, as both protocols are identical for this case.}
    \label{fig:times}
\end{figure}

To verify this result, we have calculated $t_\text{A}$ for the simulations done for Figs.~\ref{fig:bound} and \ref{fig:boundmitigated}. We show the results comparing both times in Fig.~\ref{fig:times}. The times for implementing the error mitigated circuits are larger than the ones in which the indeterminate forms are removed from the equation.

\bibliography{main.bib}

\end{document}